\documentclass[12pt,preprint]{aastex}

\begin{document}

\title{A Comparison Between Lucky Imaging and Speckle Stabilization for Astronomical Imaging}

\author{
Mark~Keremedjiev\altaffilmark{1},
and Stephen~S.~Eikenberry\altaffilmark{1},\altaffilmark{2}
\email{msk@astro.ufl.edu}
}
\altaffiltext{1}{Department of Astronomy, University of Florida, 
Gainesville, FL 32611}
\altaffiltext{2}{University of Florida Research Foundation Professor of Astronomy, Gainesville, FL 32611}

\begin{abstract}

The new technique of Speckle Stabilization \citep[][]{E08,K08,K10} has great potential to provide optical imaging data at the highest angular resolutions from the ground. While Speckle Stabilization was initially conceived for integral field spectroscopic analyses, the technique shares many similarities with speckle imaging (specifically shift-and-add and Lucky Imaging). Therefore, it is worth comparing the two for imaging applications. We have modeled observations on a 2.5-meter class telescope to assess the strengths and weaknesses of the two techniques.  While the differences are relatively minor, we find that Speckle Stabilization is a viable competitor to current Lucky Imaging systems. Specifically, we find that Speckle Stabilization is 3.35 times more efficient (where efficiency is defined as signal-to-noise per observing interval) than shift-and-add and able to detect targets 1.42 magnitudes fainter when using a standard system.  If we employ a high-speed shutter to compare to Lucky Imaging at 1\% image selection, Speckle Stabilization is 1.28 times more efficient and 0.31 magnitudes more sensitive.  However, when we incorporate potential modifications to Lucky Imaging systems we find the advantages are significantly mitigated and even reversed in the 1\% frame selection cases.  In particular, we find that in the limiting case of Optimal Lucky Imaging, that is zero read noise {\it and} photon counting, we find Lucky Imaging is 1.80 times more efficient and 0.96 magnitudes more sensitive than Speckle Stabilization.  For the cases in between, we find there is a gradation in advantages to the different techniques depending on target magnitude, fraction of frames used and system modifications. Overall, however, we find that the real strength of Lucky Imaging is in observations of the brightest targets at all frame selection levels and in observations of faint targets at the 1\% level.  For targets in the middle, we find that Speckle Stabilization regularly achieves higher $S/N$ ratios.

\end{abstract}

\keywords{} 


\section{Introduction}


Speckle Stabilization (SS) is a new technique that has the potential to achieve diffraction-limited spatial resolutions in the optical regime from ground-based telescopes \citep[][]{E08,K08,K10}. The technique is based on a relatively simple idea. On short timescales, atmospheric turbulence gets frozen out into a speckle pattern.  Each one of these moderate-to-low Strehl speckles is at the diffraction-limited spatial resolution of the telescope. By tracking the speckle patterns in real time with a high speed camera (such as an electron-multiplying CCD, EMCCD), it is possible to identify the location of the brightest speckle and stabilize it onto a traditional ``low-speed'' science camera through the use of a fast steering mirror. Over time the system tracks and stabilizes bright speckles as they appear and the result is a diffraction-limited core surrounded by a diffuse halo of scattered light. This was demonstrated through simulation by \citet{K08}. In many ways, SS is simply a real-time implementation of the shift-and-add (SAA) technique developed by \citet{BC80} who showed that by stacking speckle images on top of one another based on the location of the best quality speckle, one could extract much higher spatial resolutions. 

Recently, on-sky tests by \citet{K10} using a SS proof-of-concept instrument have produced z' images with spatial resolutions of $\approx 3\lambda /D$ on the 4.2-meter William Herschel Telescope. The instrument consists of a Andor DU-860 EMCCD for speckle sensing, an Optics-in-Motion fast steering mirror for guiding and an SBIG ST-237 as a science camera with light picked off via a beam splitter. Using this instrument, they were also able to resolve the components of the binary star system WDS 14411+1344 which were blended in the seeing-limit. This instrument is the first implementation of the Stabilized sPeckle Integral Field Spectrograph (SPIFS) envisioned by \citet{E08}. SPIFS will be a system capable of achieving diffraction-limited angular resolutions at optical wavelengths. What is unique about this instrument over other techniques is that it would exploit these high angular resolutions with an integral field spectrograph (IFS).

An IFS behind a SS system highlights a primary advantage of SS over Lucky Imaging and SAA: long exposures are possible in the science field. This enables both faint target imaging and spectroscopy. As a result, SPIFS could be coupled to an 8- or 10-meter class telescope and produce resolutions as fine as 10 milliarcseconds complete with associated spectral information (although at low Strehl values $\approx2\%$). As diffraction-limited angular resolutions are currently elusive in the optical from the ground for IFU work, SPIFS will be able to probe into astrophysical structures at unprecedented resolutions. SS will enable supermassive black holes to be measured in many more galaxies via the calcium triplet line, young stellar objects will have more of their structure probed, and even dense stellar fields will be spectroscopically classified. 

A particular example of the SPIFS potential was given in \citet{E08} describing how SPIFS will be able to resolve the optical jets produced by the micro-quasar SS 433. In the simulations, the jets were assumed to be 30 mas from the source as per the model developed by \citet{E01}.  At this distance, a broad band image of the system would merely reveal elongation. However, when integral field spectroscopy is performed, the distinct locations of the jet and counter-jet become apparent. These observations would be invaluable to further our understanding of jet astrophysics as well as accretion processes. The ability to access spectroscopic information is one of the reasons why SPIFS is being developed with 8 or 10-meter class telescopes in mind and is the primary advantage of a speckle-stabilization system over any other type of speckle imaging system. 

In this paper, we also examine Lucky Imaging-- a descendant of the SAA technique. For Lucky Imaging, the driving principle is that random fluctuations in atmospheric turbulence will occasionally result in high-quality, high-Strehl images. This technique was first proposed by \citet{F78} who calculated the fraction of time one would expect high Strehl images. Because these fluctuations are not necessarily long lived, a Lucky Imager needs to take thousands of images at high speeds and selects only the highest quality ones for analysis \citep[][]{Bald01,T02,Law06,Law07}. The resulting PSF from Lucky imaging observations is similar to a stabilized speckle PSF. That is, there is a diffuse halo with a distinct, sharp core. The primary difference is the fraction of light present in the diffraction-limited core, i.e. the Strehl ratio. The key point, however, is that both Lucky imaging and SS produce \emph{the same} spatial resolutions in the core. 

Lucky Imaging has produced interesting science in the fields of high-resolution companion searches where it has helped constrain the binary fraction of M dwarfs and M subdwarfs (Law et al. 2006, Lodieu et al. 2009, Bergfors et al. 2010) and has also been helpful in determining if exoplanet systems also have multiple stellar components \citep[][]{D09}. Lucky Imaging, by itself, acquires a usable fraction ($\approx 5-20\%$) of high-Strehl speckle images up to telescopes 2.5-m in diameter, however much beyond that size the fraction of time when high-Strehl images are present is effectively zero \citep[][]{F67,F78,Law07,S09}. Because of this drop off in high-Strehl images,  Lucky Imaging produces data nearly identical to SAA on large telescope. As a way to increase the potential of Lucky Imaging on large telescopes, \citet{Law09} have put a Lucky Imager behind an adaptive optics system at Palomar and were able to acquire near diffraction-limited spatial resolutions at Strehl ratios of $\approx 0.10$.

The purpose of this paper, however, is to compare Speckle Stabilization to SAA and Lucky Imaging purely in imaging mode. We note that there are inherent differences between SS and speckle techniques and our goal is to determine if these differences result in a significant advantage to SS. One of the differences we are interested in is the number of read outs each technique requires. SS only needs to read out the science camera a few times during observations, whereas Lucky Imaging has to read out thousands of times. Even though the read noise on the EMCCDs employed by speckle teams is typically very low (on the order of $0.2e^-$/pix), it is still present and we hypothesize it can become a significant noise term when observing faint targets. Another significant term we think might have an effect on sensitivity is the additional $\sqrt{2}$ noise term present in images taken with an EMCCD. This stochastic noise is introduced in the readout process and can only be worked around in photon-counting mode.

Therefore, we present simulations to determine if a Speckle Stabilization imager has the capability to overcome these features of Lucky Imaging systems in a way that is advantageous to an observer.


\section{Methods}


\subsection{Speckle Stabilization Simulations}
To compare Speckle Stabilization to SAA and Lucky Imaging, we need to ensure our estimations of the various parameters are accurate. We chose to conduct our comparison with a simulated 2.5-meter telescope. This particular size was chosen because at larger apertures, while Lucky Imaging and SS both produce the same spatial resolutions in the core, the Strehl ratios and usable fraction of speckle images for Lucky Imaging decreases dramatically. As a result, a 2.5-meter telescope is one where Lucky Imaging is highly effective in terms of Strehl ratios and resolution gain. 

Strehl ratios of 0.3 are common in Lucky Imaging observations in a 2.5-m telescope, but we need to have a good understanding of the Strehl ratio of SS at this aperture size. To simulate the Speckle Stabilization capabilities, we have carried out a range of simulations using model atmospheres based on the turbulence spectrum of \citet{K41}. The adaptive optics group at the Jet Propulsion Laboratory has used a similar algorithm extensively and has verified its accuracy in comparison with actual performance results with the Palomar Adaptive Optics system (PALAO).

Details of the simulation description are given in \citet{K08}, but here we outline the general procedure. We began by defining a phase map which we will use to characterize a Kolmogorov turbulence screen projected onto the pupil of the telescope. This means we require that the turbulent wavefront amplitude as a function of spatial frequency ($k$) and the Fried parameter ($r_0$) goes as $\Phi(k)\propto r_0k^{-11/3}$. We then generate another array of random phase values ranging from $-\pi$ to $+\pi$ and apply them to our phase screen. The resulting array is a random wavefront phase map $\phi(x,y)$ sampled from a Kolmogorov power spectrum.

We then define an amplitude mask $A(x,y)$ representing a telescope pupil where we assumed the secondary mirror obscures 1/3 of the primary. The final wavefront map is then given by $\Psi (x,y)=A(x,y)*\exp{(-i\phi (x,y))}$. We then determine the resulting point spread function according to $PSF(\alpha,\beta)=|FFT(\Psi(x,y))|^2$, where $\alpha$ and $\beta$ are the angular positions at the telescope focal plane and $FFT$ is a Fast-Fourier Transform. The result is a stellar speckle pattern.

Because the physical scale of our simulations is defined to be wavelength independent, a correction was applied to compare different wavelengths. The output of our simulations assumed that the diffraction-limited FWHM was 3-pixels across for each $\lambda$. Therefore, since the diffraction limit scales as $\lambda$, we needed to expand our images to match physical results. We assumed an $r_0$ for $\lambda=0.7\mu m$ and set that as our baseline size. For any image corresponding to a longer wavelength, we stretched the image out from the center by a factor of $\lambda/0.7\mu m$ using a linear interpolation.

Using this code we produced 500 distinct speckle patterns (frames) to form an SDSS i'=0 mag star as produced by a 2.5-meter telescope. Each speckle pattern was sampled at five different wavelengths (0.70$\mu m$, 0.75$\mu m$, 0.80$\mu m$, 0.85$\mu m$, and 0.90$\mu m$) to account for the broadband nature of the imaging. To simulate Speckle Stabilization, we found the ``best'' speckle in each frame using a 2D cross-correlation between the speckle pattern and an ideal PSF.  This ideal PSF was produced using the same code but with no turbulence applied . We then shifted the images according to the location of the best speckle and stacked them. To contrast to seeing-limited observations, we simply added the frames on top of one another with no shifting whatsoever. We present the results of the simulations in Fig. 1.  There we demonstrate that the stabilized image core has a FWHM similar to the diffraction-limited case. Analysis reveals that the SS PSF is only 6\% wider than the diffraction-limited case.

Strehl was measured by comparing to an ideal PSF also produced by the code. We used to FWHM of the ideal PSF to define an aperture where pixels intensity would be measured. The Strehl was measured as the intensity within the aperture measured for the SS image divided by the intensity measured within the aperture of the ideal image.  This gives a Strehl ratio of 0.085 which is similar to the \citet{T02} measured Strehl ratios of $\approx 0.06$ in Lucky Imaging observations when using 100\% frame selection (SAA). Therefore we find these simulations confirm that the SS PSF is quite similar to the SAA PSFs produced in real observations.  

\begin{figure} 
\begin{center}
\resizebox{\hsize}{!}{\includegraphics[angle=0]{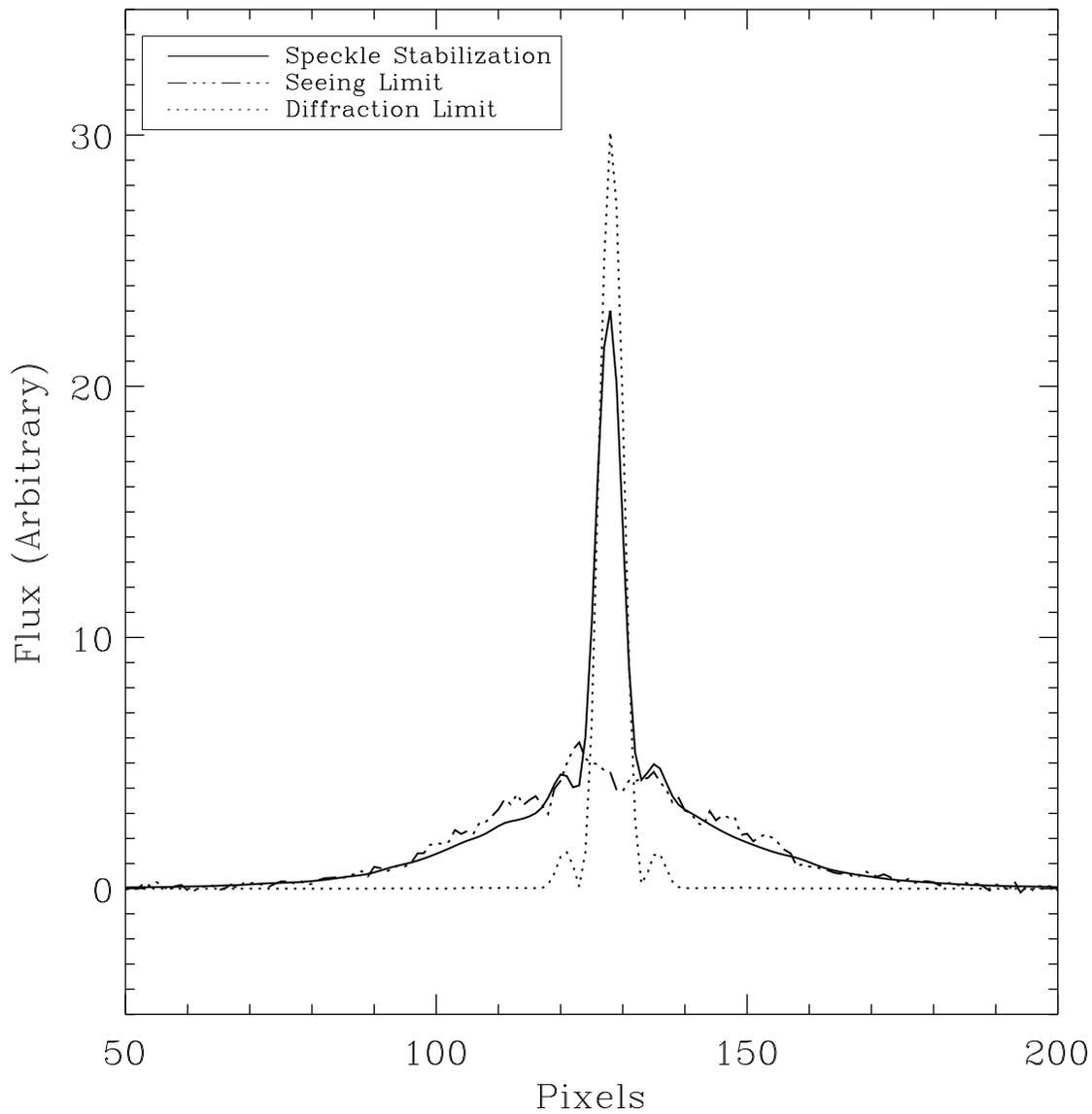}}
\caption{Cross sections of the PSF produced by Speckle Stabilization, the seeing limit, and the diffraction limit. The diffraction-limited image has been arbitrarily scaled to demonstrate that the SS PSF has a similar FWHM. Note that the seeing-limit has a similar FWHM to the extended halo of the speckle-stabilization image.
\label{fig1}}
\end{center}
\end{figure}

\subsection{Comparison Between Methods}
In order to compare SS to SAA and Lucky Imaging, we will examine theoretical signal-to-noise ratios as a function of target magnitude. Since we wish to make the comparison as realistic and direct as possible, we use parameters from current Lucky Imaging systems like LuckyCam, AstraLux and FastCam \citep[][]{Law07,H08,O08} and use a comparable standard CCD for Speckle Stabilization. Current Lucky Imagers use EMCCDs of sizes around 512 x 512 pixels$^2$, so we will model a standard CCD of this size for the SS as well. We note that the EMCCD used for the actual speckle stabilization would be a 128 x 128 pixel$^2$ CCD and is thus capable of much faster readouts.

To perform the comparisons between the two techniques, we employ the following equation for Lucky Imaging:

\begin{equation}
S/N=\frac{\omega \alpha \beta f T_{Exp}}{\sqrt{2(\omega \alpha \beta \delta f T_{Exp}+\omega n_{pix}(\alpha f_{sky}T_{Exp}+DT_{Exp}+n_{Exp}r^2))}}
\end{equation}

Here $f$ is the number of photons per second received from a star of a given magnitude at a 2.5-meter telescope in the Sloan \emph{i'} filter. The fraction of photons that are transmitted through the optics of the system is $\alpha$ which we assume  to equal 0.5. The fraction of photons that are present in a diffraction-limited core is given by $\delta$. We assume that there are perfect optics producing an Airy pattern-- as a result, we assume $\delta=0.838$. The Strehl ratio is given by $\beta$. For Lucky Imaging we assume a Strehl of 0.30 for this telescope diameter at 1\% frame selection and Strehl=0.20 for 10\% frame selection \citep[][]{Bald01,T02,Law07,Bald08}. For SAA (100\% frame selection) we assume Strehl of 0.085-- the Strehl calculated from our simulations in \S2.1 as SAA has a similar PSF to SS. The total time on source is $T_{Exp}$ and is determined by the number of exposures, $n_{Exp}$ times exposure length $t_{Exp}$. We assume $n_{Exp}=50000$ and $t_{Exp}=0.030$ seconds meaning the total time spent ``on source'' is 1500 sec. An exposure time of 30 milliseconds is used because the coherence length of the atmosphere, generally given by $t_0\approx r_0/v$ (where $r_0$ is the Fried parameter and $v$ is the bulk wind velocity of the dominant turbulent layer Kern et al. 2000), is on the order of 30 milliseconds. There is no time assumed for the read out as all current Lucky Imaging systems use frame-transfer EMCCDs and a 30ms is greater than or equal to the read out time for a 512 x 512 pixel$^2$ EMCCD \citep[][]{Law07,H08,O08}.

We also use the term $\omega$, which is the fraction of near diffraction-limited images kept, which we assume to be 0.01 and 0.10 for Lucky Imaging cases and 1.0 for SAA. This choice of $\omega=0.01$ and $0.10$ for Lucky Imaging was made because we wanted to examine the case of best-possible Strehl improvement as well as more typical Lucky Imaging parameters. 

In the noise term, the $\sqrt{2}$ is a result of the stochastic processes involved in the readout of an EMCCD. We assume Nyquist sampling so that the central core of the source covers four pixels, meaning $n_{pix}=4$. We use $f_{sky}$ to include the effects of sky background. For this model, we assume dark time and use $i'=19.9$ $mag/arcsec^2$ for the sky level. We use $D$ as a measure of dark current and apopt of value of $0.002$ $e^-/pix/sec$. 

Finally $r$ is the read noise is assumed to be $0.2$ $e^-/pix$. We chose $0.2$ $e^-/pix$ because lab tests of our own Andor iXon DU 860 at EM Gain of 300 converged to this value as does the documentation that came with our camera. We use EM Gain=300 for two reasons. The first is that this is a typical value used by non-cryogenic EMCCDs for Lucky Imaging (Femenia, B. private correspondence). The second reason is that the Solis program written by Andor for the camera will not allow the user to select gain values higher than this number and strongly encourages using EM Gain lower than 300. We note, however, that through the use of the Andor Software Development Kit it is possible to attain higher EM Gain levels through custom written code and discuss the ramifications of higher levels in \S3.1.

A similar equation is used for Speckle Stabilization:

\begin{equation}
S/N=\frac{\omega \alpha \beta \delta f T_{Exp}}{\sqrt{\omega \alpha \beta \delta f T_{Exp}+\omega n_{pix}(\alpha f_{sky}T_{Exp}+DT_{Exp}+n_{Exp,SS}r^2)}}
\end{equation}

The equation is mostly the same as Eqn. 1, but there are several subtle differences in the parameters. In the SS case, $\omega=1.0$ since no images are thrown out. We also use a Strehl ratio of 0.085 for SS as determined from our simulations above in \S2.1. A read noise of $5$ $e^-/pix$ was assumed for the science channel, but the $\sqrt{2}$ term disappears for the SS calculations.

To directly compare these two methods, we need to ensure that the time spent by the telescope per target is the same. By doing this, we define efficiency as the signal-to-noise (\emph{S/N}) achieved per observing interval. Where the observing interval is defined as $T_{Exp,SS}=t_{Exp,SS}*n_{Exp,SS}$. For the SS observations, the number of exposures is calculated by $n_{Exp,SS}=T_{Exp,Speckle}/(t_{Exp,SS}+t_{readout,SS})$. In this case, $t_{Exp,SS}$ is the time it takes to saturate an image or $T_{Exp,Speckle}$, whichever is smaller.  We assume a pixel is ``saturated'' when there have been 64000 photon events. We have fixed $t_{readout,SS}$ at 1 second as this is the typical time it takes to read out a 512 x 512 pixels$^2$ standard CCD.

In its current form, SS acts like SAA as both have similar PSFs and Strehl ratios. Comparisons between these two techniques are valid, but SS compared to the higher Strehl Lucky Imaging cases make less sense. While the low-contrast images are still of high scientific value, one way to boost the Strehl ratios of Speckle Stabilization systems is the installation of a high-speed shutter in front of the science camera \citep[][]{K10}. With a shutter fast enough to open and close at $>100$ Hz, a speckle stabilization system would be able ensure only high-Strehl data find their way onto the detector. In this way, a SS would act like a real-time Lucky Imaging selection algorithm. Therefore, we also include models which include a high-speed shutter to compare to the lower frame-selection, higher-Strehl Lucky Imaging data. We refer to this technique as Speckle Stabilization + Shutter (SS+S). The only modifications to Eqn. (2) needed to characterize a shutter-based system are that we change $\omega=0.01$ and $0.10$ to reflect the fraction of Lucky Imaging images typically selected, but conversely we would also get the higher Strehls, $b=0.30$ and $0.20$. Although for dark current, we would still use the full observing interval as the detector continues to accumulate dark charge with the shutter closed.

One further note we wish to address is that of guide stars. For these models, we assume that there is a bright guide star near enough that there is little or no degradation in the image quality of the analysis star. We make this assumption because we are only interested in the theoretical limits of these two techniques and to probe the faint magnitudes of these tests requires a bright, nearby guide star for both techniques.


\section{Results and Discussion}


We calculated the results of equations (1) and (2) for stars from \emph{i'}=0 to \emph{i'}=30 mag and present the results in Fig. 2. There are six curves in the top part of the figure. The red lines represent Lucky Imaging and SAA while the black lines denote SS and SS+S. A blue dashed line is plotted at $S/N=3$-- a common detection threshold. In the bottom part of the figure we present the efficiency ratio, defined as $(S/N_{SS}) / (S/N_{Lucky})$, as a function of magnitude. Values less than 1 denote cases where Lucky Imaging has higher signal-to-noise per observing interval and values greater than 1 denote cases where SS has higher signal-to-noise per observing interval. A blue dashed line is also plotted to show where the two techniques are equal.

Here we demonstrate that the technique of speckle stabilization has advantages and disadvantages as compared to Lucky Imaging. For most of the models, SS and Lucky Imaging have the same shape and are offset by a factor of $\approx \sqrt{2}$. This is to be expected as in the non-saturated, non-photon starved middle ground, the biggest difference is the additional $\sqrt{2}$ term in the noise for Lucky Imaging. Where the noticeable differences occur are in the bright and faint ends. At the bright end, we see that SS and SS+S have bright cut-off points. This occurs because SS requires integration times long enough to allow the system to stabilize speckles. The current SS prototype operates at 100 Hz and future plans are in the works to operate at 500 Hz \citep[][]{K10}. Therefore, we will use 100 Hz as a baseline and require 10 cycles (with the shutter open) to have elapsed to get the minimum necessary corrections. This means we define the minimum exposure time as $\omega T_{Exp,SS}=0.1$. Using this convention, we see that SAA can observe targets brighter than $i'=6.87$, Lucky Imaging at 10\% frame selection can observe targets brighter than $i'=7.80$ and Lucky Imaging at 1\% frame selection can observe targets brighter than $i'=8.24$ whereas the corresponding SS and SS+S techniques cannot. 

\begin{figure} 
\begin{center}
{\includegraphics[angle=0,scale=0.7]{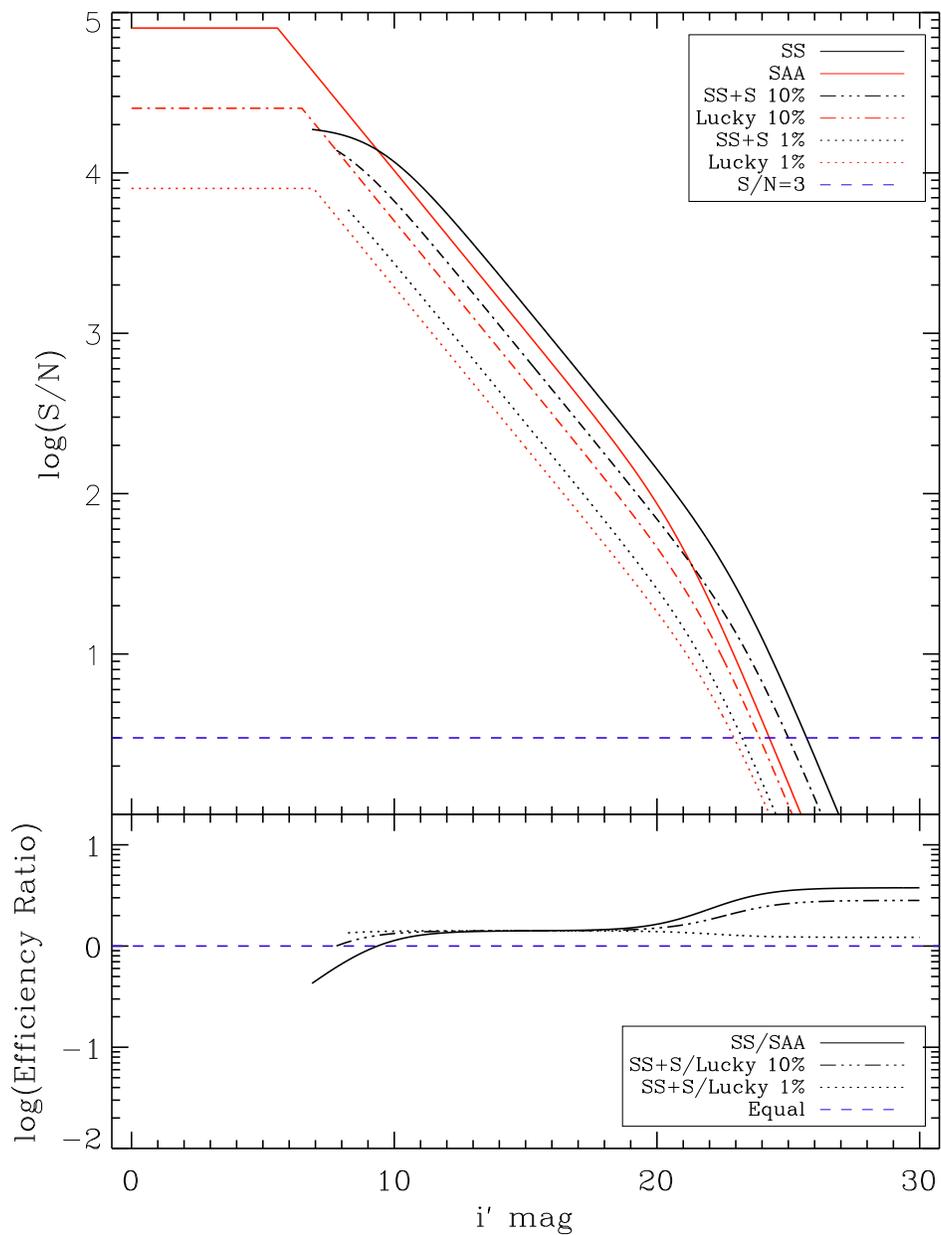}}
\caption{Comparison between SS and SS+S with SAA and Lucky Imaging. Also plotted is a blue line denoting $S/N=3$, a common detection limit. On the lower part of the plot are the ratio between \emph{S/N$_{SS}$} and \emph{S/N$_{Lucky}$}. These ratios are the efficiency ratios. Also plotted on the lower plot in blue is a line denoting an efficiency ratio of 1-- where both systems perform the same.
\label{fig2}}
\end{center}
\end{figure}

At the fainter end, advantages to SS and SS+S become heightened. We see that speckle stabilization techniques have higher efficiencies and sensitivities than Lucky Imaging (again where efficiency is defined as \emph{S/N} per observing interval). Sensitivity is the difference in magnitudes at the $3\sigma$ detection level. Specifically, SS is 3.35 times more efficient and 1.42 magnitudes more sensitive than SAA at the SAA detection limit. SS+S at 10\% selection is 2.40 times more efficient and 1.10 magnitudes more sensitive than corresponding Lucky Imaging and SS+S at 1\% selection is 1.28 times more efficient and 0.31 magnitudes more sensitive than the corresponding Lucky Images. While these values are not excessively large, they are interesting. The fact that the difference becomes more pronounced with higher frame selection points to the fact that both sky background and read noise are more significant in Lucky Imaging than in Speckle Stabilization. 

These results are fairly robust. Changing most of the shared parameters like $f$, $\alpha$, telescope diameter, etc have little impact on the overall results-- doing so only shifts the functions vertically or horizontally. Modifying the saturation limits also has little impact as it only affects where and how large the plateau exists for brighter targets. The biggest factors are the Strehl ratios and the read noise terms.

\subsection{Read Noise Limit}

From the above analysis, we see that SS is able to produce higher \emph{S/N} values and reach fainter magnitudes.  However, this advantage is not particularly large. As a result, we chose to look at modifications that can be made to Lucky Imaging and SAA systems to improve their efficiency. The first modification we examine is adjusting the read noise. One of the primary advantages of L3CCDs and EMCCDs over conventional CCDs is the they have very little read noise. This attribute is a result of the electron multiplication process where very high gain values (10-10000) are involved. This means that a real photon strike results in a very large measurement-- far greater than the noise of the electronics. While this fact gives these CCDs very high sensitivities, it is also the reason for their additional $\sqrt{2}$ noise term.

In the cases where exceptionally high gain values ($g>>300$) are employed, it is possible to reduce the read noise to effectively zero (although this generally requires the use of cryogenically cooled cameras as dark current can have dramatic effects). Therefore, we decided to compare SS to Lucky Imaging in the limiting case where read noise equals zero. We present the results in Fig. 3. For the 1\% frame selection Lucky Imaging data, there does not appear to be any truncation all the way down to the detection limit. In the SS+S 1\% data, a truncation around i'=20 still occurs when sky background begins to affect the data. These models show that no read noise \emph{Lucky Imaging} is actually a factor of 1.28 times more efficient than SS+S at the SS+S detection limit and is 0.41 magnitudes more sensitive. 

Looking at the SS vs. SAA case, however, a slightly different picture emerges. Here the read noise in the SS data is less prevalent and sky background appears to affect both models. SS manages to be 1.36 times more efficient and 0.36 magnitudes more sensitive, but we note that these vales are about a factor of two less than with standard read noise in the EMCCD. The case of 10\% frame selection actually is somewhat of a hybrid of these two extremes as SS+S and Lucky Imaging models move toward convergence at the faint end.

Therefore, we see that reducing the read noise to zero removes much of the advantage of Speckle Stabilization techniques over their partner speckle imaging techniques. In fact, at 1\% image selection, Lucky Imaging is both more sensitive and efficient than SS+S.

\begin{figure} 
\begin{center}
{\includegraphics[angle=0,scale=0.7]{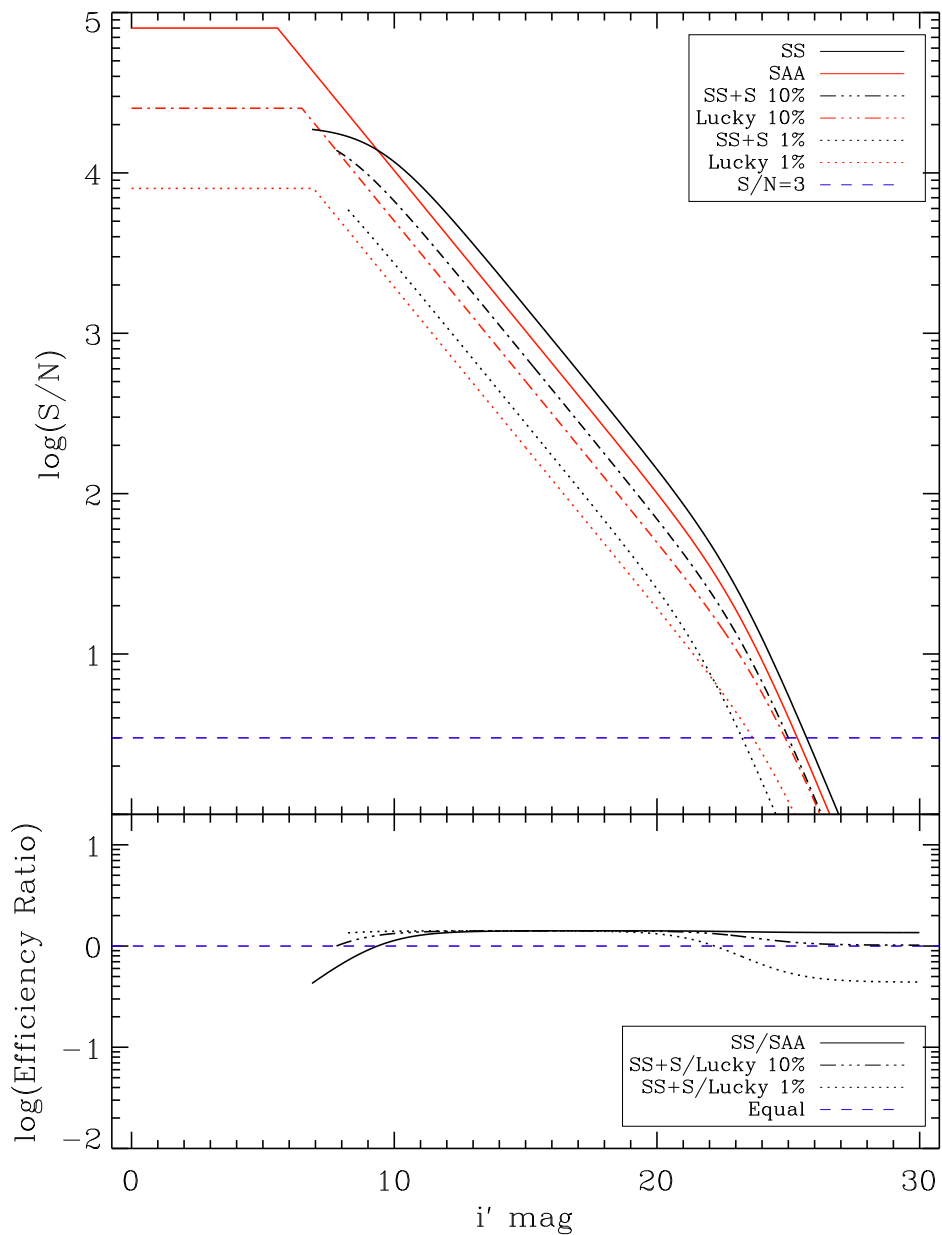}}
\caption{Comparison between Speckle Stabilization with and without shutter and no-read-noise SAA and Lucky Imaging. In the case of Lucky Imaging, there is no truncation in the function presented. Also plotted is a line denoting $S/N=3$, a common detection limit. The lower part of the plot is the same as Fig. 2, but in this case, the dotted black line is the ratio of the \emph{S/N} SS+Shutter to Lucky Imaging.
\label{fig3}}
\end{center}
\end{figure}

\subsection{Photon Counting}

Here we address photon counting mode as a way to mitigate some of the advantage of Speckle Stabilization. With data collected from an EMCCD, it is possible to do photon counting in post-processing whereby a pixel either measures one photon or none. This scenario is advantageous because it overcomes the additional $\sqrt{2}$ shot noise usually associated with electron multiplication. It is only useful, however, in cases where there is extremely little flux per exposure. Once more than one photon per pixel per exposure occurs, the advantage is mitigated. 

To model the photon counting case, we assumed that when there was $<1$ photon/pixel/frame, we would switch on photon counting and drop the extra $\sqrt{2}$ term in the noise. For cases where the counts were higher than this, we assumed standard Lucky Imaging analysis. 

We present the photon counting case in Fig. 4. The location where photon counting switches on is immediately apparent in the figure a this is the location where the \emph{S/N} of the speckle techniques jumps in a discontinuous fashion.  When this occurs, Lucky Imaging data exactly match the SS and SS+S data for a few magnitudes. In the 1\% selection case, Lucky Imaging actually tracks SS+S all the way down to the detection limit and even slightly out-performs it with a factor of 1.12 boost in efficiency and an increase of 0.14 magnitudes in sensitivity.

\begin{figure} 
\begin{center}
{\includegraphics[angle=0,scale=0.7]{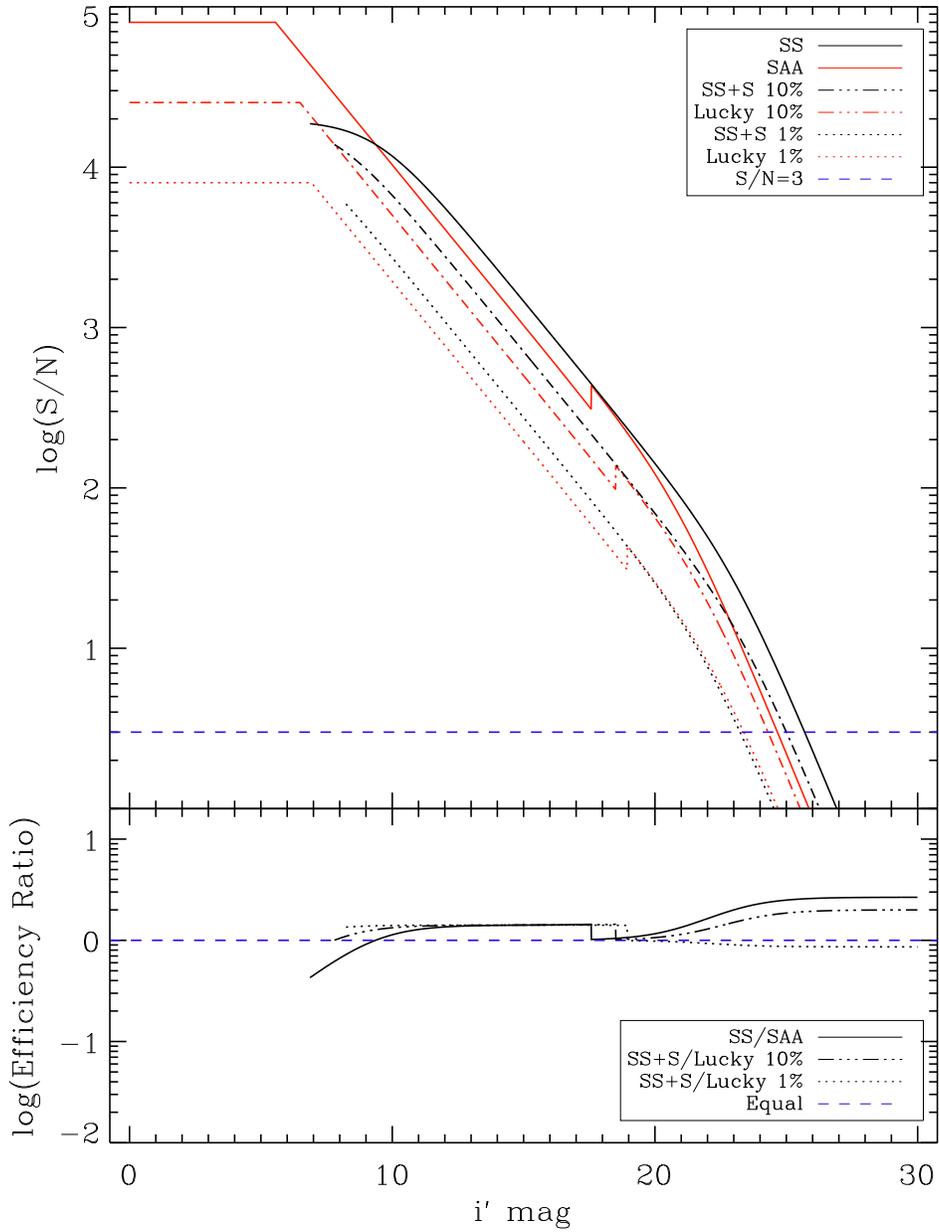}}
\caption{Comparison between techniques in the case of photon counting. The discontinuous jumps in the speckle imaging data are due to the point where photon counting becomes active. Also plotted is a line denoting $S/N=3$, a common detection limit. The lower part of the plot is the same as Fig. 2.
\label{fig4}}
\end{center}
\end{figure}

The 10\% and 100\% frame selection, however, are not as efficient or as sensitive as their corresponding SS+S and SS models. This is largely due to the fact that the read noise term is still present and becomes dominant at these fainter magnitudes. Specifically, SS+S is 1.76 times more efficient and 0.70 magnitudes more sensitive than Lucky Imaging at 10\% frame selection and SS is 2.44 times more efficient and 1.04 magnitudes more sensitive than SAA.

Overall, this shows that photon counting is a valuable way to get more information out of Lucky Imaging data, and can be used to observe fainter targets with much higher sensitivity.

\subsection{Optimal Lucky Imaging}

From the previous sections, it is clear that there are circumstances in which Lucky Imaging and SAA are able to mitigate the positive gains of Speckle Stabilization.  As such, we discuss the case of an Optimal Lucky Imager (OLI).  To maximize the effectiveness of a Lucky Imaging system, the ideal design would be an extremely low read noise system with photon counting performed for fainter targets. We examine this case in Fig. 5 where read noise is zero and photon counting enabled.

Our models reveal that an OLI is both more efficient and sensitive for Lucky Imaging than SS+S and has nearly equivalent performance when comparing SAA to SS. In particular we find OLI is 1.80 times more efficient at the detection limit and 0.96 magnitudes more sensitive than SS+S with 1\% frame selection and 1.29 times more efficient and 0.32 magnitudes more sensitive with 10\% frame selection. However, on average, SS+S is still marginally more efficient at brighter magnitudes. SS and SAA have nearly the same properties in terms of sensitivity and efficiency. Overall, this means that with fairly few modifications, it is possible to optimize a Lucky Imaging system so that it is maximally efficient and sensitive.

Furthermore, one issue we have not addressed in our simulations is an inherent advantage to Lucky Imaging: because all the analysis is done off-line, in post processing image selection algorithms can be tuned to maximize \emph{S/N} and Strehl ratios. The observer can also decide at what point to enable photon counting in post processing as well. This gives Lucky Imaging much more flexibility with respect to data products and is part of the optimization process.

\begin{figure} 
\begin{center}
{\includegraphics[angle=0,scale=0.7]{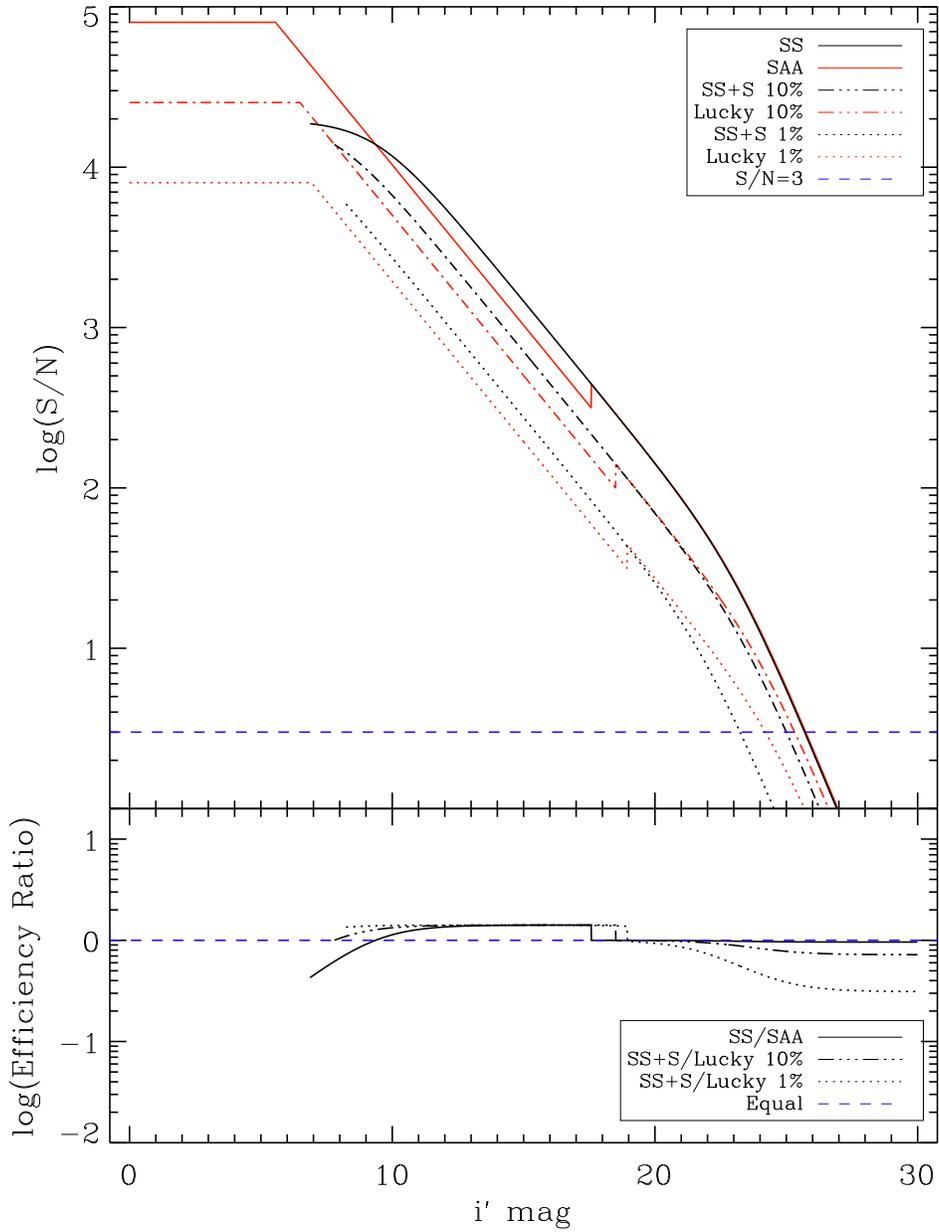}}
\caption{Comparison between speckle stabilization and OLI. The discontinuous jumps in the speckle imaging data are due to the point where photon counting becomes active. Also plotted is a line denoting $S/N=3$, a common detection limit. The lower part of the plot is the same as Fig. 2.
\label{fig5}}
\end{center}
\end{figure}

\subsection{1024x1024 pixel$^2$ Detectors}

To be thorough, we look at one final case to show the potential of Speckle Stabilization. This final case increases the size of the detectors to 1024 x 1024 pixels$^2$ enabling wide-field imaging. At this size, the advantages of a Speckle Stabilization system becomes heightened. This is because the read out times for a 1024 x 1024 pixels$^2$ Lucky Imaging system are currently quite long at $\approx 100$ms. This means even in frame transfer mode, 30ms integrations still require 100ms to read out. To highlight this point, in Fig. 8 we demonstrate the comparison between an OLI system and a SS system both with 1024 x 1024 pixels$^2$ detectors. For the SS detector, we assume a read out time of 4 seconds.

\begin{figure} 
\begin{center}
{\includegraphics[angle=0,scale=0.7]{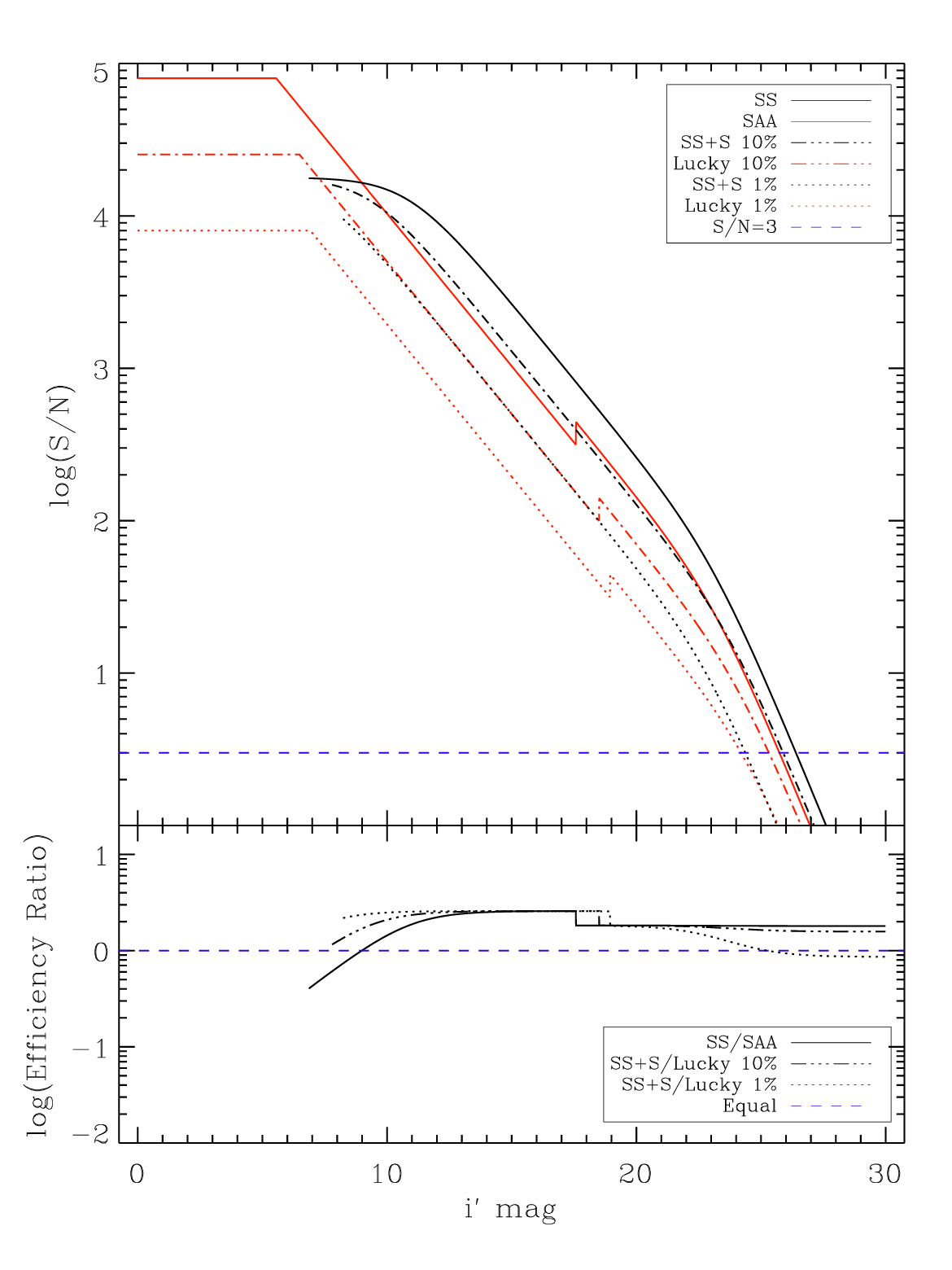}}
\caption{Comparison between speckle stabilization and OLI for 1024x1024 pixel$^2$ detectors. The discontinuous jumps in the speckle imaging data are due to the point where photon counting becomes active. Also plotted is a line denoting $S/N=3$, a common detection limit. The lower part of the plot is the same as Fig. 2.
\label{fig6}}
\end{center}
\end{figure}

In this case, because the readout times are so long for an EMCCD of this size, Speckle Stabilization has an advantage in both sensitivity and efficiency for both the shutter and non-shutter cases. When comparing SS+S to Lucky Imaging, we find that SS+S is 1.15 times more efficient at the faintest magnitude and 0.17 magnitudes more sensitive with an average efficiency ratio of 2.22 with 1\% frame selection. For 10\% frame selection, these values are 1.63 times more efficient at the Lucky Imaging detection threshold and 0.57 more sensitive. In the case of SS and SAA, we find SS is 1.81 times more efficient and 0.66 magnitudes more sensitive.

We note that being able to use a detector of this size has great scientific potential. At 30 mas pixel sampling, a 1024 x 1024 pixels$^2$ detector would have a FOV of 30 arcsec. While this kind of FOV is a bit larger than the isoplanatic patch, \citet{K08} have shown that speckle stabilization is effective out to offsets as large as 30 arcseconds and would be of high scientific value.

\begin{table} 
\caption{{Summary of simulation results. E is defined as the efficiency ratio at the limiting magnitude of the Lucky Imaging system. $\Delta S$ is the increase in sensitivity of SS over Lucky Imaging. }\label{tab1}}
\resizebox{\textwidth}{!}{
\begin{tabular}{llllllllll}
 & SS 100\% & & & SS+S 10\% & & & SS+S 1\% & & \\
Simulation & Eff. & $\Delta S$ (mag) & $<$Eff.$>$ & Eff. & $\Delta S$ (mag) & $<$Eff.$>$ & Eff. & $\Delta S$ (mag) & $<$Eff.$>$  \\
Standard        & 3.35  & 1.42 & 1.54 & 2.40  & 1.10 & 1.26 & 1.28  & 0.31 & 1.39 \\
No Read Noise   & 1.36  & 0.36 & 1.28 & 1.11  & 0.12 & 1.34 & 0.787 &-0.41 & 1.33 \\
Photon Counting & 2.44  & 1.04 & 1.33 & 1.76  & 0.70 & 1.33 & 0.896 &-0.14 & 1.28 \\
OLI             & 0.964 &-0.04 & 1.10 & 0.778 &-0.32 & 1.20 & 0.556 &-0.96 & 1.24 \\
1024 x 1024     & 1.81  & 0.66 & 1.86 & 1.63  & 0.57 & 2.11 & 1.15  & 0.17 & 2.22 \\
\end{tabular}}
\end{table}

\section{Conclusions}

We find that Speckle Stabilization is a viable competitor to current Lucky Imaging systems when used solely for imaging in certain circumstances. The results from all of our models and simulations are presented in Table 1. 

Both SS and Lucky Imaging have their own strengths. In general, SS is more sensitive and efficient than speckle imaging at the faintest magnitudes with normal frame selection meaning it could be well-employed for faint object work. Additionally, for most mid-magnitude targets SS and SS+S is a factor of $\sqrt{2}$ times more efficient owing to the lack of the additional noise term.

As a counter, there are several cases in which speckle techniques are superior. We find that the Lucky Imaging techniques are the only way to observe the brightest stars with any usable \emph{S/N}. Our work has also revealed that simple modifications to traditional Lucky Imaging systems can greatly improve their performance and completely mitigate the advantage of a SS system at the faint end. The most effective alterations are approaching zero read noise and using photon counting techniques beyond a particular threshold. When these features are implemented, we found the Lucky Imaging was both more sensitive and efficient than Speckle Stabilization at the faint end.

We highlight again that the main advantage to Speckle Stabilization is long exposures for IFU work, but this paper has revealed SS is also useful from an imaging perspective. Overall the differences are fairly minor between the output products, but as telescope time is a valuable commodity it is useful to have instruments in place that are able to perform efficient observations. We find that speckle stabilization is one way to achieve this aim, but similar goals can be met by modifying existing Lucky Imaging systems. While SS is still in its infancy, instruments like SPIFS will be able to reveal some of the potential of this technique and help solve outstanding issues in astrophysics.


\acknowledgments 
Mark would like to thank Joe Carson for his discussions during the first SPIFS prototype run. Simply asking the fundamental question: ``how does SPIFS compare Lucky for imaging?'' led to this paper. He would also like to thank Lauren Young for her keen eye and valuable comments in editing this manuscript. Both authors would like to thank the anonymous referee for closely reading the draft and providing comments which greatly improved the paper. This work was supported by NSF SGER 0917758.


\end{document}